\documentclass[prb,amsmath,superscriptaddress,showpacs,showkeys,twocolumn]{revtex4}
\usepackage{amssymb,amsmath}   
\usepackage[dvips]{graphicx}   
\usepackage{verbatim}   
\usepackage{color}      
\usepackage{subfigure}  
\usepackage{hyperref}   
\usepackage{gensymb}
\usepackage{epstopdf}
\usepackage{natbib}
\usepackage{enumerate}
\usepackage{mathbbol}

\begin{document}

\def\omegav{\mbox{\boldmath{$\omega$}}}
\def\sigmav{\mbox{\boldmath{$\sigma$}}}
\def\kappav{\mbox{\boldmath{$\kappa$}}}
\def\piv{\mbox{\boldmath{$\pi$}}}
\def\Wv{\mbox{\boldmath{$W$}}}
\def\Bv{\mbox{\boldmath{$B$}}}
\def\nv{\mbox{\boldmath{$n$}}}
\def\pv{\mbox{\boldmath{$p$}}}
\def\rv{\mbox{\boldmath{$r$}}}
\def\sv{\mbox{\boldmath{$s$}}}
\def\gv{\mbox{\boldmath{$g$}}}
\def\Ev{\mbox{\boldmath{$E$}}}
\def\Kv{\mbox{\boldmath{$K$}}}
\def\Av{\mbox{\boldmath{$A$}}}

\title{Robustness of spin filtering against current leakage in a Rashba-Dresselhaus-Aharonov-Bohm interferometer}
\author{Shlomi Matityahu}
\affiliation{Department of Physics, Ben-Gurion University, Beer
Sheva 84105, Israel}
\author{Amnon Aharony}
\email{aaharony@bgu.ac.il} \affiliation{Department of Physics,
Ben-Gurion University, Beer Sheva 84105, Israel} \affiliation{Ilse
Katz Center for Meso- and Nano-Scale Science and Technology,
Ben-Gurion University, Beer Sheva 84105, Israel}
\affiliation{Raymond and Beverly Sackler School of Physics and
Astronomy, Tel Aviv University, Tel Aviv 69978, Israel}
\author{Ora Entin-Wohlman}
\affiliation{Department of Physics, Ben-Gurion University, Beer
Sheva 84105, Israel} \affiliation{Ilse Katz Center for Meso- and
Nano-Scale Science and Technology, Ben-Gurion University, Beer
Sheva 84105, Israel} \affiliation{Raymond and Beverly Sackler
School of Physics and Astronomy, Tel Aviv University, Tel Aviv
69978, Israel}
\author{Shingo Katsumoto}
\affiliation{Institute for Solid State Physics, University of
Tokyo, Kashiwa, Chiba 277-8581, Japan}
\date{\today}
\begin{abstract}
In an earlier paper [Phys. Rev. B {\bf 84}, 035323 (2011)], we
proposed a spin filter which was based on a diamond-like
interferometer, subject to both an Aharonov-Bohm flux and (Rashba
and Dresselhaus) spin-orbit interactions. Here we show that the
full polarization of the outgoing electron spins remains the same
even when one allows leakage of electrons from the branches of the
interferometer. Once the gate voltage on one of the branches is
tuned to achieve an effective symmetry between them, this
polarization can be controlled by the electric and/or magnetic
fields which determine the spin-orbit interaction strength and the
Aharonov-Bohm flux.

\end{abstract}

 \maketitle
\section{Introduction}
\label{Intro} Conventional technology usually exploits the
electric charge of the electron. In the last two decades, a new
technology has emerged called spintronics, which involves the
active control and manipulation of the spin degree of freedom in
condensed matter devices. \cite{WSA01,ZI04,BSD10} Adding the spin
degree of freedom to the conventional charge-based technology has
the potential advantages of multifunctionality, longer decoherence
times and lengths, increased data processing speed, decreased
electric power consumption, and increased integration densities
compared with conventional semiconductor devices. Besides being
useful in contemporary technology, spintronics may also contribute
to the field of quantum computation and quantum
information.\cite{Nielsen&Chuang} Spin-1/2 is a natural candidate
for the quantum bit (qubit) realization. In a spin-based quantum
computer, the information is contained in the unit vector along
which the spin is polarized. Writing and reading information on a
spin qubit is thus equivalent to polarizing the spin along a
specific direction and identifying the direction along which the
spin is polarized, respectively. We distinguish between two
different realizations of spin qubits, namely static and mobile
qubits. In static qubit realizations the information transfer is
accomplished by transferring the state of the qubit, rather than
the qubit itself. For instance, if the qubits are represented by
the spins of electrons localized on a quantum
dot,\cite{LD98,KFLH06,PM08} then the information is transferred
along a chain of quantum dots via exchange interactions between
neighboring dots. With mobile qubits,\cite{BCHW00,PAE04} the qubit
itself is moved around the quantum circuit, carrying the
information from one point to another. A major advantage of those
over static ones is the ability to manipulate mobile qubits by
static electric and magnetic fields in predefined regions rather
than by expensive high-frequency electromagnetic
pulses.\cite{HT03} Here we consider mobile qubits. In Ref.
\onlinecite{BCHW00} it has been proposed to implement a system of
mobile spin qubits in a two-dimensional electron gas (2DEG) by
using surface acoustic waves (SAW) that capture single electrons
in their potential minima and drag them through a parallel
connection of $N$ quantum one-dimensional channels. A single
quantum computation is performed by the $N$ electrons in a single
SAW minimum as they are being dragged through a pattern of
magnetic and nonmagnetic quantum gates. The feasibility of this
setup was demonstrated for two parallel channels ($N=2$) by
Ebbecke \textit{et al.}.\cite{EJ00}

For a quantum computation it is necessary to provide qubits in
pure states.\cite{Nielsen&Chuang} Hence, a major aim of
spintronics is to build mesoscopic spin valves (or spin filters),
which polarize the spins going through them along tunable
directions. At first glance, the easiest way to construct such
devices is by using ferromagnets that inject and/or collect
polarized electrons. However, the connection of ferromagnets to
semiconductors is inefficient, due to a large impedance mismatch
between them.\cite{ZI04,SG00} A completely different approach is
to use spintronic devices that do not involve ferromagnetism at
all.\cite{KY04,ADD09} Here we discuss such filters that avoid
ferromagnets.

Recently, several groups proposed spin filters based on a single
loop, subject to both electric and magnetic fields perpendicular
to the plane of the loop.\cite{CR06,HN07,AA11} An important
question that has not been addressed in those studies is whether
spin filtering is robust against current leakage, for example due
to quantum tunnelling out of the loop. Furthermore, in a practical
device, with the application of finite source-drain bias voltage,
the wires which connect the four dots are inevitably charged up if
they are electrostatically isolated. To avoid such probably
unfavorable effects, one needs to ground them, which leads to
current leakage. In this paper we examine this question in the
context of the diamond loop subject to both an Aharonov-Bohm (AB)
flux\cite{AY59} and a spin-orbit interaction (SOI), discussed in
our previous paper (see Fig. \ref{fig:diamond
interferometer1}).\cite{AA11} There, we have shown that with a
certain symmetry between the two branches of the diamond, and with
appropriate tuning of the electric and magnetic fields (or of the
diamond shape), this device serves as a perfect spin filter as
well as a spin analyzer.

\begin{figure}[ht]
\centering
\includegraphics[width=0.5\textwidth]{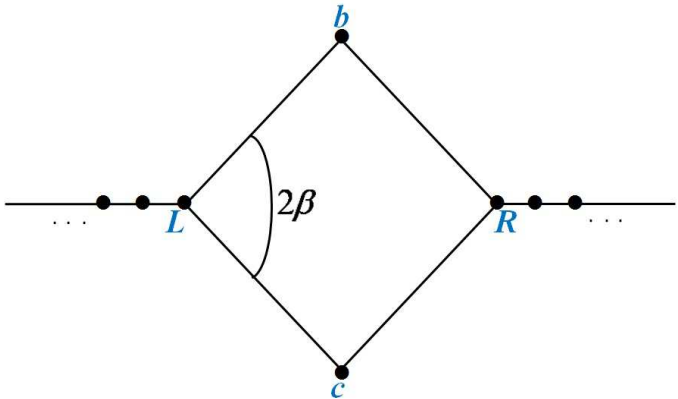}
\caption{\label{fig:diamond interferometer1} The lossless diamond.
The diamond is penetrated by a magnetic flux $\Phi$, and its edges
(of length $L$) are subject to spin-orbit interactions.}
\end{figure}

In the presence of an electric field, free electrons experience
the well-known SOI in vacuum,\cite{Sakurai}
\begin{align}
\label{eq:Spin-orbit in vacuum}
&\mathcal{H}^{}_{SO}=\Lambda\sigmav\cdot\left[\pv\times\nabla
V(\rv)\right].
\end{align}
Here, $\Lambda=\hbar/\left(2m^{}_{0}c\right)^{2}$ ($m^{}_{0}$ is
the mass of a free electron and $c$ is the speed of light in
vacuum), $\pv$ is the electron momentum, $V(\rv)$ is the electric
scalar potential, and the Pauli matrices $\sigmav$ are related to
the electron spin via $\sv=\hbar\sigmav/2$. In a 2DEG formed in
mesoscopic structures, made of narrow-gap semiconductor
heterostructures, the SOI is modified due to an external applied
potential (e.g.\ gate voltage) and a periodic lattice
potential.\cite{Winkler} The final result can often be recast as
an effective SOI Hamiltonian, of the general form
$\mathcal{H}^{}_{SO}=(\hbar k^{}_{SO}/m)(\piv\cdot\sigmav)$, where
$k^{}_{SO}$ characterizes the SOI strength, $\piv$ is a linear
combination of the electron momentum components $p^{}_{x}$ and
$p^{}_{y}$ and $m$ is the effective mass, usually much smaller
than $m^{}_{0}$. The related energy scale can be larger than that
of Eq. \eqref{eq:Spin-orbit in vacuum} by as much as six orders of
magnitude.

Two special cases of the linear (in the momentum) SOI should be
emphasized, namely the Rashba SOI\cite{Rashba} and the Dresselhaus
SOI.\cite{Dresselhaus} The Rashba SOI is a result of a confining
potential well which is asymmetric under space inversion. For an
electric field $\Ev=-\nabla V$ in the $z$ direction, this SOI has
the form
\begin{align}
\label{eq:Rashba Hamiltonian} &\mathcal{H}^{}_{R}=\frac{\hbar
k^{}_{R}}{m}\left(p^{}_{y}\sigma^{}_{x}-p^{}_{x}\sigma^{}_{y}\right).
\end{align}
The coefficient $k^{}_{R}$ depends on the magnitude of $\Ev$, and
can be controlled by a gate voltage, as shown in several
experiments.\cite{NJ97,KT02,KM06,BT06,LD12} The Dresselhaus SOI
results from a lattice potential which lacks inversion symmetry.
For a 2DEG this SOI is given by
\begin{align}
\label{eq:Dresselhaus Hamiltonian} &\mathcal{H}^{}_{D}=\frac{\hbar
k^{}_{D}}{m}\left(p^{}_{x}\sigma^{}_{x}-p^{}_{y}\sigma^{}_{y}\right),
\end{align}
where $k^{}_{D}$ usually depends on the crystal structure and only
weakly (if at all) on the external field. When a spin moves in the
presence of these SOIs a distance $L$ in the direction of the unit
vector $\hat{\gv}$, its spinor $|\chi\rangle$ transforms into
$|\chi'\rangle=U|\chi\rangle$, with the unitary spin rotation
matrix $U=e^{i\Kv\cdot\sigmav}$.\cite{OY92,BD04} Here, the vector
$\Kv$ is
\begin{align}
\label{eq:Rotation matrix1}
&\Kv=\alpha^{}_{R}\left(-g^{}_{y},g^{}_{x},0\right)+\alpha^{}_{D}\left(-g^{}_{x},g^{}_{y},0\right),
\end{align}
with the dimensionless coefficients $\alpha^{}_{R,D}\equiv
k^{}_{R,D}L$. Below we use the unitary matrix $U$ and the
parameters $\alpha^{}_{R,D}$ to characterize the hopping between
adjacent bonds in the presence of SOI.

In addition to the SOI related phase, electrons also gain an AB
phase $\phi$ in the presence of a magnetic flux $\Phi$ penetrating
the loop.\cite{AY59} When an electron goes around a loop, its wave
function gains an AB phase $\phi\equiv2\pi\Phi/\Phi^{}_{0}$, where
$\Phi^{}_{0}=hc/e$ is the flux quantum ($e$ is the electron
charge).\cite{comment1} The combined effect of the SOI and the AB
flux is to transform the spinor $|\chi\rangle$ of an electron that
goes around a loop into $|\chi'\rangle=u|\chi\rangle$, where the
unitary matrix $u$ is of the form
\begin{align}
\label{eq:Rotation matrix2}
&u=u^{}_{\text{AB}}u^{}_{\text{SOI}}=e^{i\phi+i\omegav\cdot\sigmav}.
\end{align}
Here $u^{}_{\text{AB}}=e^{i\phi}{\bf 1}$ (${\bf 1}$ is the
$2\times 2$ unit matrix) is the diagonal transformation matrix due
to the AB flux and $u^{}_{\text{SOI}}=e^{i\omegav\cdot\sigmav}$ is
the transformation matrix due to the SOI. The latter is a product
of matrices of the form $e^{i\Kv\cdot\sigmav}$ discussed above,
each coming from the local SOI on a segment of the
loop.\cite{AA11}

In the present paper we study the sensitivity of the spin filter
to leakage of currents from the branches of the interferometer. To
model this leakage, we generalize the diamond interferometer shown
in Fig. \ref{fig:diamond interferometer1} by allowing for an
arbitrary number of tight-binding sites along each edge of the
diamond. Each site is connected to a one-dimensional lead, which
allows only an outgoing current to an absorbing reservoir (see
Fig. \ref{fig:diamond interferometer2} and a detailed description
below).\cite{AA02} We calculate the spin-dependent transmission
through the diamond and thus generalize the results of Ref.
\onlinecite{AA11} to include the effects of the leakage. We show
that the diamond interferometer may still serve as a perfect spin
filter and spin analyzer, even in the presence of current leakage.
With slight modifications, the requirements for the symmetry
between the two branches and for a specific relation between the
AB flux and the SOI strength, derived in Ref. \onlinecite{AA11},
are preserved. There we presented conditions for achieving this
symmetry independent of the electrons' energy. These conditions
were quite difficult to be realized. Here we show that under
linear-response conditions, when all the electrons have the same
Fermi energy, it is relatively easy to achieve this symmetry by
tuning a single gate voltage on one of the interferometer
branches. Furthermore, we show that once the filtering conditions
are obeyed, the blocked polarization and the polarization of the
outgoing electrons are not affected by the current leakage, being
the same as for the lossless diamond.

The outline of the paper is as follows: in Sec. \ref{Sec 1} we
first define the model and present a general calculation of the
transmission through the diamond, valid for any internal structure
of the two one-dimensional paths (Sec. \ref{Sec 1A}), and then
find the conditions for full filtering (Sec. \ref{Sec 1B}). The
differences and similarities between the lossless and lossy
diamonds are analyzed. The results are discussed and summarized in
Sec. \ref{Summary}.
\section{The Lossy diamond}
\label{Sec 1}
\subsection{Details of the model}
\label{Sec 1A} 
\begin{figure}[ht]
\centering
\includegraphics[width=0.5\textwidth]{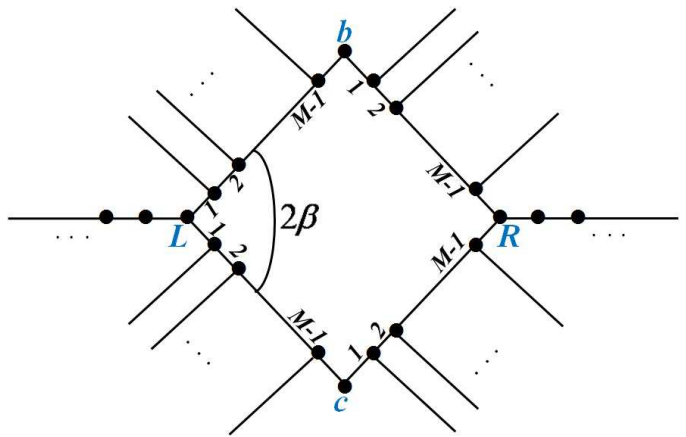}
\caption{\label{fig:diamond interferometer2} The lossy diamond.
The diamond is penetrated by a magnetic flux $\Phi$, and its edges
(of length $L$) are subject to spin-orbit interaction. Electrons
can tunnel out of the interferometer from sites $n=1, \ldots, M-1$
on edge $uv$ ($uv=Lb, Lc, cR, bR$) through absorbing channels.}
\end{figure}

Let us consider the scattering of spin-1/2
electrons by a lossy diamond with arbitrary SOI and AB flux, as
shown in Fig. \ref{fig:diamond interferometer2}. Each edge $uv$
($uv=Lb, Lc, cR, bR$) of the diamond consists of $M+1$
tight-binding sites with lattice constant $a$, labelled from left
to right. The length of each edge is $L=Ma$ and we allow for
arbitrary opening angle, $2\beta$ of the diamond. The sites on the
corners are labelled as $L$, $R$, $b$ and $c$ (Fig.
\ref{fig:diamond interferometer2}) and are characterized by site
energies $\epsilon^{}_{0,L}$, $\epsilon^{}_{0,R}$,
$\epsilon^{}_{0,b}$ and $\epsilon^{}_{0,c}$, respectively. The
sites $n=1, \ldots, M-1$ on each edge $uv$ have zero site energies
$\epsilon^{}_{0}=0$. To allow for a possible current leakage, each
of these sites is connected to an absorbing channel, modelled as a
one-dimensional tight-binding chain with site energies
$\epsilon^{}_{0}=0$ and free of SOI. Electrons can tunnel out of
the interferometer through these absorbing channels. The hopping
amplitude on the first bond on each absorbing channel is
$J^{}_{x,uv}$, while the other bonds have a hopping amplitude $j$.
The diamond is connected at sites $L$ and $R$ to two leads with
site energies $\epsilon^{}_{0}=0$ and hopping amplitudes $j$.

The tight-binding Schr\"{o}dinger equations for the spinors
$|\psi^{uv}_{n}\rangle$ at sites $n=1, \ldots, M-1$ on edge $uv$
of the diamond are
\begin{align}
\label{eq:Equation for chain}
&\epsilon|\psi^{uv}_{n}\rangle=-J^{}_{uv}\left(U^{}_{uv}|\psi^{uv}_{n-1}\rangle+U^{\dagger}_{uv}|\psi^{uv}_{n+1}\rangle\right)-J^{}_{x,uv}|\Psi^{uv}_{0}\rangle,
\end{align}
where $U^{}_{uv}$ is a $2\times 2$ unitary matrix, $J^{}_{uv}$ is
a hopping amplitude for bonds on edge $uv$, and
$|\Psi^{uv}_{0}\rangle$ is the spinor at the first site of the
absorbing channel, connected to one site on edge $uv$ of the
diamond. The corresponding equations for the spinors
$|\Psi^{uv}_{n}\rangle$ at sites $n=0, 1, \ldots$ on each of the
absorbing channels are
\begin{align}
\label{eq:Equation for teeth}
&\epsilon|\Psi^{uv}_{n}\rangle=-j\left(|\Psi^{uv}_{n-1}\rangle+|\Psi^{uv}_{n+1}\rangle\right),
\qquad n\geq 1, \nonumber\\
&\epsilon|\Psi^{uv}_{0}\rangle=-j|\Psi^{uv}_{1}\rangle-J^{*}_{x,uv}|\psi^{uv}_{n}\rangle.
\end{align}
Assuming only outgoing waves on the absorbing channels, with
spinors $|\Psi^{uv}_{n}\rangle=t^{}_{uv}|\chi^{uv}\rangle
e^{ikna}$, and energy $\epsilon=-2j\cos\left(ka\right)$, one can
eliminate the spinor
$|\Psi^{uv}_{0}\rangle=t^{}_{uv}|\chi^{uv}\rangle$ from Eqs.
\eqref{eq:Equation for teeth}.\cite{AA02} Substitution into Eqs.
\eqref{eq:Equation for chain} yields
\begin{align}
\label{eq:Equation for chain2}
&\left(\epsilon-\widetilde{\epsilon}^{}_{uv}\right)|\psi^{uv}_{n}\rangle=-J^{}_{uv}\left(U^{}_{uv}|\psi^{uv}_{n-1}\rangle+U^{\dagger}_{uv}|\psi^{uv}_{n+1}\rangle\right),
\end{align}
with the site self energy $\widetilde{\epsilon}^{}_{uv}$ given by
\begin{align}
\label{eq:Effective site energy}
&\widetilde{\epsilon}^{}_{uv}=-\frac{|J^{}_{x,uv}|^{2}e^{ika}}{j}.
\end{align}
The tight-binding Schr\"{o}dinger equations \eqref{eq:Equation for
chain2} are easily solved with the transformation
\begin{align}
\label{eq:spinors transformation}
|\psi^{uv}_{n}\rangle=U^{n}_{uv}|\varphi^{uv}_{n}\rangle,
\end{align}
by which Eqs. \eqref{eq:Equation for chain2} read
\begin{align}
\label{eq:Equation for chain3}
&\left(\epsilon-\widetilde{\epsilon}^{}_{uv}\right)|\varphi^{uv}_{n}\rangle=-J^{}_{uv}\left(|\varphi^{uv}_{n-1}\rangle+|\varphi^{uv}_{n+1}\rangle\right).
\end{align}
Therefore, in terms of the transformed spinors
$|\varphi^{uv}_{n}\rangle$, the tight-binding equations
\eqref{eq:Equation for chain3} include neither the AB flux nor the
SOI. The solution in terms of the spinors at the diamond corners
is thus\cite{AA09}
\begin{align}
\label{eq:phi spinors}
&|\varphi^{uv}_{n}\rangle=\frac{\sin\left[k^{}_{uv}\left(M-n\right)a\right]|\psi^{}_{u}\rangle+\sin\left(k^{}_{uv}na\right)U^{-M}_{uv}|\psi^{}_{v}\rangle}{\sin\left(k^{}_{uv}Ma\right)},
\end{align}
where the wave vector $k^{}_{uv}$ satisfies the equation
\begin{align}
\label{eq:wave vectors}
&\epsilon-\widetilde{\epsilon}^{}_{uv}=-2J^{}_{uv}\cos\left(k^{}_{uv}a\right).
\end{align}
For $J^{}_{x,uv}\neq0$ the solution of Eq. \eqref{eq:wave vectors}
yields a complex wave vector, which implies an exponential decay
of the propagating waves, due to a leakage of part of the
electrons out of the interferometer into the absorbing channels.

The tight-binding Schr\"{o}dinger equations for the spinors at the
diamond corners are
\begin{align}
\label{eq:corners equations}
&\left(\epsilon-\epsilon^{}_{0,L}\right)|\psi^{}_{L}\rangle=-J^{}_{Lb}U^{\dagger}_{Lb}|\psi^{Lb}_{1}\rangle-J^{}_{Lc}U^{\dagger}_{Lc}|\psi^{Lc}_{1}\rangle-j|\psi^{L}_{-1}\rangle, \nonumber\\
&\left(\epsilon-\epsilon^{}_{0,b}\right)|\psi^{}_{b}\rangle=-J^{}_{bR}U^{\dagger}_{bR}|\psi^{bR}_{1}\rangle-J^{}_{Lb}U^{}_{Lb}|\psi^{Lb}_{M-1}\rangle, \nonumber\\
&\left(\epsilon-\epsilon^{}_{0,c}\right)|\psi^{}_{c}\rangle=-J^{}_{cR}U^{\dagger}_{cR}|\psi^{cR}_{1}\rangle-J^{}_{Lc}U^{}_{Lc}|\psi^{Lc}_{M-1}\rangle, \nonumber\\
&\left(\epsilon-\epsilon^{}_{0,R}\right)|\psi^{}_{R}\rangle=-J^{}_{cR}U^{}_{cR}|\psi^{cR}_{M-1}\rangle-J^{}_{bR}U^{}_{bR}|\psi^{bR}_{M-1}\rangle \nonumber\\
&\qquad\qquad\qquad\qquad\!\!-j|\psi^{R}_{M+1}\rangle,
\end{align}
with $|\psi^{L}_{n}\rangle$ ($n\leq0$) and $|\psi^{R}_{n}\rangle$
($n\geq M$) being the spinors at the sites on the left and right
leads, respectively. By using the transformation \eqref{eq:spinors
transformation} together with Eq. \eqref{eq:phi spinors} for the
transformed spinors, Eqs. \eqref{eq:corners equations} take the
form
\begin{align}
\label{eq:corners equations2}
&\left(\epsilon-\epsilon^{}_{L}\right)|\psi^{}_{L}\rangle=-J''^{}_{Lb}U^{-M}_{Lb}|\psi^{}_{b}\rangle-J''^{}_{Lc}U^{-M}_{Lc}|\psi^{}_{c}\rangle-j|\psi^{L}_{-1}\rangle, \nonumber\\
&\left(\epsilon-\epsilon^{}_{b}\right)|\psi^{}_{b}\rangle=-J''^{}_{bR}U^{-M}_{bR}|\psi^{}_{R}\rangle-J''^{}_{Lb}U^{M}_{Lb}|\psi^{}_{L}\rangle, \nonumber\\
&\left(\epsilon-\epsilon^{}_{c}\right)|\psi^{}_{c}\rangle=-J''^{}_{cR}U^{-M}_{cR}|\psi^{}_{R}\rangle-J''^{}_{Lc}U^{M}_{Lc}|\psi^{}_{L}\rangle, \nonumber\\
&\left(\epsilon-\epsilon^{}_{R}\right)|\psi^{}_{R}\rangle=-J''^{}_{cR}U^{M}_{cR}|\psi^{}_{c}\rangle-J''^{}_{bR}U^{M}_{bR}|\psi^{}_{b}\rangle \nonumber\\
&\qquad\qquad\qquad\quad-j|\psi^{R}_{M+1}\rangle,
\end{align}
with the site self energies
\begin{align}
\label{eq:effective site energies}
&\epsilon^{}_{L}=\epsilon^{}_{0,L}-J'^{}_{Lb}-J'^{}_{Lc}, \nonumber\\
&\epsilon^{}_{b}=\epsilon^{}_{0,b}-J'^{}_{bR}-J'^{}_{Lb}, \nonumber\\
&\epsilon^{}_{c}=\epsilon^{}_{0,c}-J'^{}_{Lc}-J'^{}_{cR}, \nonumber\\
&\epsilon^{}_{R}=\epsilon^{}_{0,R}-J'^{}_{cR}-J'^{}_{bR},
\end{align}
and with
\begin{align}
\label{eq:effective_hopping1}&J'^{}_{uv}=J^{}_{uv}\frac{\sin\left[k^{}_{uv}\left(M-1\right)a\right]}{\sin\left(k^{}_{uv}Ma\right)}, \\
\label{eq:effective_hopping2}&J''^{}_{uv}=J^{}_{uv}\frac{\sin\left(k^{}_{uv}a\right)}{\sin\left(k^{}_{uv}Ma\right)}.
\end{align}
Equations \eqref{eq:corners equations2} are analogous to Eqs. (7)
of Ref. \onlinecite{AA11}. All the modifications caused by the
current leakage are embodied in the site self energies
$\epsilon^{}_{L}$, $\epsilon^{}_{b}$, $\epsilon^{}_{c}$, and
$\epsilon^{}_{R}$ [Eqs. \eqref{eq:effective site energies} and
\eqref{eq:effective_hopping1}] and the complex effective hopping
amplitudes $J''^{}_{Lb}$, $J''^{}_{Lc}$, $J''^{}_{cR}$,
$J''^{}_{bR}$ [Eqs. \eqref{eq:effective_hopping2}].\cite{comment2}
Elimination of $|\psi^{}_{b}\rangle$ and $|\psi^{}_{c}\rangle$
from Eqs. \eqref{eq:corners equations2} yields
\begin{align}
\label{eq:corners equations3}
&\left(\epsilon-y^{}_{L}\right)|\psi^{}_{L}\rangle=\Wv^{}_{RL}|\psi^{}_{R}\rangle-j|\psi^{L}_{-1}\rangle, \nonumber\\
&\left(\epsilon-y^{}_{R}\right)|\psi^{}_{R}\rangle=\Wv^{}_{LR}|\psi^{}_{L}\rangle-j|\psi^{R}_{M+1}\rangle,
\end{align}
where
\begin{align}
\label{eq:y's}
&y^{}_{L}=\epsilon^{}_{L}+\frac{J''^{2}_{Lb}}{\epsilon-\epsilon^{}_{b}}+\frac{J''^{2}_{Lc}}{\epsilon-\epsilon^{}_{c}}, \nonumber\\
&y^{}_{R}=\epsilon^{}_{R}+\frac{J''^{2}_{bR}}{\epsilon-\epsilon^{}_{b}}+\frac{J''^{2}_{cR}}{\epsilon-\epsilon^{}_{c}},
\end{align}
\begin{align}
\label{eq:W matrices}
&\Wv^{}_{LR}=\gamma^{}_{b}U^{}_{b}+\gamma^{}_{c}U^{}_{c}, \nonumber\\
&\Wv^{}_{RL}=\gamma^{}_{b}U^{\dagger}_{b}+\gamma^{}_{c}U^{\dagger}_{c},
\end{align}
with the complex coefficients
\begin{align}
\label{eq:gamma}
&\gamma^{}_{b}=\frac{J''^{}_{Lb}J''^{}_{bR}}{\epsilon-\epsilon^{}_{b}}, \nonumber\\
&\gamma^{}_{c}=\frac{J''^{}_{Lc}J''^{}_{cR}}{\epsilon-\epsilon^{}_{c}},
\end{align}
and the unitary matrices
\begin{align}
\label{eq:U matrices}
&U^{}_{b}=U^{M}_{bR}U^{M}_{Lb}, \nonumber\\
&U^{}_{c}=U^{M}_{cR}U^{M}_{Lc}.
\end{align}
The matrices $U^{}_{b}$ and $U^{}_{c}$ correspond to transitions
from left to right through the upper and lower paths,
respectively.

Consider a wave coming from the left, namely
\begin{align}
\label{eq:left wave}
&|\psi^{L}_{n}\rangle=|\chi^{}_{in}\rangle e^{ikna}+r|\chi^{}_{r}\rangle e^{-ikna}, \qquad n\leq 0, \nonumber\\
&|\psi^{R}_{n}\rangle=t|\chi^{}_{t}\rangle
e^{ik\left(n-M\right)a}, \qquad n\geq M,
\end{align}
where $|\chi^{}_{in}\rangle$, $|\chi^{}_{r}\rangle$ and
$|\chi^{}_{t}\rangle$ are the incoming, reflected and transmitted
normalized spinors, respectively, with the corresponding
reflection and transmission amplitudes $r$ and $t$. The
transmission and reflection amplitude matrices are defined by the
relations
\begin{align}
\label{eq:Transmission and Reflection}
&t|\chi^{}_{t}\rangle\equiv\mathcal{T}|\chi^{}_{in}\rangle, \qquad
r|\chi^{}_{r}\rangle\equiv\mathcal{R}|\chi^{}_{in}\rangle.
\end{align}
To calculate these matrices, we substitute Eqs. \eqref{eq:left
wave} into Eqs. \eqref{eq:corners equations3}. This
yields\cite{AA11}
\begin{align}
\label{eq:Transmission}&\mathcal{T}=2ij\sin\left(ka\right)\Wv^{}_{LR}\left(Y\textbf{1}-\Wv^{}_{RL}\Wv^{}_{LR}\right)^{-1}, \\
\label{eq:Reflection}&\mathcal{R}=-\textbf{1}-2ij\sin\left(ka\right)X^{}_{R}\left(Y\textbf{1}-\Wv^{}_{RL}\Wv^{}_{LR}\right)^{-1},
\end{align}
where
\begin{align}
\label{eq:X and Y}
&X^{}_{L,R}=y^{}_{L,R}+je^{-ika}, \nonumber\\
&Y=X^{}_{L}X^{}_{R}.
\end{align}
Compared with the expressions for the transmission and reflection
amplitudes of the lossless diamond [Eqs. (14) in Ref.
\onlinecite{AA11}], one notes the following difference. In the
lossless diamond the coefficients $\gamma^{}_{b}$ and
$\gamma^{}_{c}$ [Eqs. \eqref{eq:gamma}] are real and then,
according to Eqs. \eqref{eq:W matrices}, the relation
$\Wv^{}_{RL}=\Wv^{\dagger}_{LR}$ holds. In the lossy diamond, on
the other hand, the coefficients $\gamma^{}_{b}$ and
$\gamma^{}_{c}$ are complex and
$\Wv^{}_{RL}\neq\Wv^{\dagger}_{LR}$. Hence
$\Wv^{}_{RL}\Wv^{}_{LR}$, involved in both $\mathcal{T}$ and
$\mathcal{R}$, is not an hermitian matrix as in the lossless case.
Nevertheless, in the next subsection we show that spin filtering
may still be obtained in the presence of current leakage by
appropriately tuning the gate voltages and the magnetic field.
\subsection{Filtering conditions for the lossy diamond}
\label{Sec 1B} In this section we study the properties of the
spin-dependent transmission matrix \eqref{eq:Transmission} and
write it in a form that enables us to derive the spin filtering
conditions. Consider first the matrix $\Wv^{}_{RL}\Wv^{}_{LR}$.
Using Eqs. \eqref{eq:W matrices}, we get
\begin{align}
\label{eq:WLRWRL2}
&\Wv^{}_{RL}\Wv^{}_{LR}=\gamma^{2}_{b}+\gamma^{2}_{c}+\gamma^{}_{b}\gamma^{}_{c}\left(u+u^{\dagger}\right),
\end{align}
where $u=U^{\dagger}_{b}U^{}_{c}$ is the unitary matrix
representing an anticlockwise hopping from site $L$ back to site
$L$ around the loop. As discussed in the introduction, the matrix
$u$ has the form $u=e^{i\phi+i\omegav\cdot\sigmav}$ and therefore
$u+u^{\dagger}=2\left(\cos\omega\cos\phi-\sin\omega\sin\phi\hat{\omegav}\cdot\sigmav\right)$.
Thus, Eq. \eqref{eq:WLRWRL2} can be written as
\begin{align}
\label{eq:WLRWRL} &\Wv^{}_{RL}\Wv^{}_{LR}=A+\Bv\cdot\sigmav,
\end{align}
with
\begin{align}
\label{A and B}
&A=\gamma^{2}_{b}+\gamma^{2}_{c}+2\gamma^{}_{b}\gamma^{}_{c}\cos\omega\cos\phi, \nonumber\\
&\Bv=2\gamma^{}_{b}\gamma^{}_{c}\sin\omega\sin\phi\:\hat{\nv}=B\hat{\nv}.
\end{align}
Here, $\hat{\nv}\equiv-\hat{\omegav}$ is a real unit vector along
the direction of $-\omegav$. Defining the eigenstates of the spin
component along an arbitrary direction $\hat{\nv}$ via
$\hat{\nv}\cdot\sigmav|\pm\hat{\nv}\rangle=\pm|\pm\hat{\nv}\rangle$,
we identify the eigenvectors of $\Wv^{}_{RL}\Wv^{}_{LR}$ as
$|\pm\hat{\nv}\rangle$, namely
\begin{align}
\label{eq:WLRWRL eigenvectors}
&\Wv^{}_{RL}\Wv^{}_{LR}|\pm\hat{\nv}\rangle=\lambda^{}_{\pm}|\pm\hat{\nv}\rangle,
\end{align}
with the corresponding eigenvalues $\lambda^{}_{\pm}$ being
\begin{align}
\label{eq:WLRWRL eigenvalues}&\lambda^{}_{\pm}=A\pm
B=\gamma^{2}_{b}+\gamma^{2}_{c}+2\gamma^{}_{b}\gamma^{}_{c}\cos\left(\phi\pm\omega\right).
\end{align}

Consider an incoming electron with its spin polarized along
$\pm\hat{\nv}$. The spinor at the output of the diamond will be
\begin{align}
\label{eq:outgoing spinor}
&t^{}_{\pm}|\chi^{\text{out}}_{\pm}\rangle=\mathcal{T}|\pm\hat{\nv}\rangle=\frac{2ij\sin\left(ka\right)}{Y-\lambda^{}_{\pm}}\Wv^{}_{LR}|\pm\hat{\nv}\rangle.
\end{align}
The transmission amplitudes $t^{}_{\pm}$ for the two opposite
polarizations are calculated from the scalar product of Eq.
\eqref{eq:outgoing spinor} with itself using the normalization
condition
$\langle\chi^{\text{out}}_{\pm}|\chi^{\text{out}}_{\pm}\rangle=1$.
Thus,
\begin{align}
\label{eq:transmission amplitudes}
&|t^{}_{\pm}|^{2}=\bigg|\frac{2ij\sin\left(ka\right)}{Y-\lambda^{}_{\pm}}\bigg|^{2}\langle\pm\hat{\nv}|\Wv^{\dagger}_{LR}\Wv^{}_{LR}|\pm\hat{\nv}\rangle.
\end{align}
To calculate the expectation value in Eq. \eqref{eq:transmission
amplitudes}, note that by using Eqs. \eqref{eq:W matrices} the
matrix $\Wv^{\dagger}_{LR}\Wv^{}_{LR}$ is found to be
\begin{align}
\label{eq:WRLWRL}
&\Wv^{\dagger}_{LR}\Wv^{}_{LR}=|\gamma^{}_{b}|^{2}+|\gamma^{}_{c}|^{2}+\left(\gamma^{*}_{b}\gamma^{}_{c}u+\gamma^{}_{b}\gamma^{*}_{c}u^{\dagger}\right).
\end{align}
Substituting $u=e^{i\phi+i\omegav\cdot\sigmav}$,
$\gamma^{}_{b}=|\gamma^{}_{b}|e^{i\delta^{}_{b}}$ and
$\gamma^{}_{c}=|\gamma^{}_{c}|e^{i\delta^{}_{c}}$ into the last
relation gives
\begin{align}
\label{eq:WRLWRL2}
&\Wv^{\dagger}_{LR}\Wv^{}_{LR}=A^{}_{LR}+\Bv^{}_{LR}\cdot\sigmav,
\end{align}
with
\begin{align}
\label{ARL and BRL}
&A^{}_{LR}=|\gamma^{}_{b}|^{2}+|\gamma^{}_{c}|^{2}+2|\gamma^{}_{b}||\gamma^{}_{c}|\cos\omega\cos\widetilde{\phi}, \nonumber\\
&\Bv^{}_{LR}=2|\gamma^{}_{b}||\gamma^{}_{c}|\sin\omega\sin\widetilde{\phi}\:\hat{\nv}=B^{}_{LR}\hat{\nv},
\end{align}
and $\widetilde{\phi}=\phi+\delta^{}_{c}-\delta^{}_{b}$. Hence,
the eigenvectors of $\Wv^{\dag}_{LR}\Wv^{}_{LR}$ are also
$|\pm\hat{\nv}\rangle$ and the corresponding eigenvalues are
\begin{align}
\label{eq:WRLWRL eigenvalues}
&\lambda^{}_{LR,\pm}=A^{}_{LR}\pm B^{}_{LR}=|\gamma^{}_{b}|^{2}+|\gamma^{}_{c}|^{2} \nonumber\\
&+2|\gamma^{}_{b}||\gamma^{}_{c}|\cos\left(\widetilde{\phi}\pm\omega\right).
\end{align}
Then, according to Eq. \eqref{eq:transmission amplitudes} the
transmission amplitudes $t^{}_{\pm}$ are
\begin{align}
\label{eq:transmission amplitudes2}
&|t^{}_{\pm}|=\frac{2j|\sin\left(ka\right)|}{|Y-\lambda^{}_{\pm}|}\sqrt{\lambda^{}_{LR,\pm}}.
\end{align}
Now let us find the direction of the transmitted spinors
$|\chi^{\text{out}}_{\pm}\rangle$. Substituting Eq.
\eqref{eq:transmission amplitudes2} into Eq. \eqref{eq:outgoing
spinor}, we get
\begin{align}
\label{eq:outgoing spinor2}
&|\chi^{\text{out}}_{\pm}\rangle=\frac{e^{-i\delta^{}_{\pm}}}{\sqrt{\lambda^{}_{LR,\pm}}}\Wv^{}_{LR}|\pm\hat{\nv}\rangle,
\end{align}
with $\delta^{}_{\pm}$ being some arbitrary phases. Using Eqs.
\eqref{eq:WLRWRL eigenvectors} and \eqref{eq:outgoing spinor2},
one finds
\begin{align}
\label{eq:outgoing spinor3}
&\Wv^{}_{LR}\Wv^{}_{RL}|\chi^{\text{out}}_{\pm}\rangle=\lambda^{}_{\pm}|\chi^{\text{out}}_{\pm}\rangle,
\end{align}
which shows that $|\chi^{\text{out}}_{\pm}\rangle$ is an
eigenstate of $\Wv^{}_{LR}\Wv^{}_{RL}$. Calculating this matrix
from Eqs. \eqref{eq:W matrices}, we find
\begin{align}
\label{eq:WRLWLR}&\Wv^{}_{LR}\Wv^{}_{RL}=\gamma^{2}_{b}+\gamma^{2}_{c}+\gamma^{}_{b}\gamma^{}_{c}\left(u'+u'^{\dagger}\right),
\end{align}
with $u'=U^{}_{b}U^{\dagger}_{c}$. The eigenvectors of the matrix
$\Wv^{}_{LR}\Wv^{}_{RL}$ correspond to some new direction
$\hat{\nv}'$, so that
$|\chi^{\text{out}}_{\pm}\rangle=|\pm\hat{\nv}'\rangle$. Therefore
we can write $\Wv^{}_{LR}$ as
\begin{align}
\label{eq:WRL}
&\Wv^{}_{LR}=e^{i\delta^{}_{-}}\sqrt{\lambda^{}_{LR,-}}|-\hat{\nv}'\rangle\langle-\hat{\nv}|+e^{i\delta^{}_{+}}\sqrt{\lambda^{}_{LR,+}}|\hat{\nv}'\rangle\langle\hat{\nv}|.
\end{align}
Similarly, Eq. \eqref{eq:outgoing spinor} implies that the
transmission amplitude matrix [Eq. \eqref{eq:Transmission}] has
the form
\begin{align}
\label{eq:Transmission2}&\mathcal{T}=t^{}_{-}|-\hat{\nv}'\rangle\langle-\hat{\nv}|+t^{}_{+}|\hat{\nv}'\rangle\langle\hat{\nv}|.
\end{align}
This form enables us to derive the explicit conditions for spin
filtering as follows.

Expanding an arbitrary incoming spinor
$|\chi^{}_{\text{in}}\rangle$ in the basis $|\pm\hat{\nv}\rangle$,
\begin{align}
\label{eq:incoming spinor}
&|\chi^{}_{\text{in}}\rangle=c^{}_{-}|-\hat{\nv}\rangle+c^{}_{+}|\hat{\nv}\rangle,
\end{align}
where
$c^{}_{\pm}\equiv\langle\pm\hat{\nv}|\chi^{}_{\text{in}}\rangle$,
the outgoing spinor is
\begin{align}
\label{eq:outgoing spinor3}
&t|\chi^{}_{\text{out}}\rangle=\mathcal{T}|\chi^{}_{\text{in}}\rangle=c^{}_{-}t^{}_{-}|-\hat{\nv}'\rangle+c^{}_{+}t^{}_{+}|\hat{\nv}'\rangle.
\end{align}
Equation \eqref{eq:outgoing spinor3} implies that the outgoing
spinor is polarized along a definite direction
$|\mp\hat{\nv}'\rangle$, provided that one of the eigenvalues
$\lambda^{}_{LR,\pm}$ vanishes. For example, if
$\lambda^{}_{LR,-}=0$, then $t^{}_{-}=0$ [Eq.
\eqref{eq:transmission amplitudes2}] and the lossy diamond then
serves as a perfect spin filter, since all outgoing electrons have
their spin polarized along $\hat{\nv}'$. Moreover, once the
parameters of the device have been appropriately tuned so that
$t^{}_{-}=0$, this device can also serve as a spin
analyzer.\cite{AA11} We emphasize that the blocked spinors,
$|\pm\hat{\nv}\rangle$, and the transmitted ones,
$|\pm\hat{\nv}'\rangle$, being the eigenvectors of the matrices
$\Wv^{}_{RL}\Wv^{}_{LR}$ [Eq. \eqref{eq:WLRWRL2}] and
$\Wv^{}_{LR}\Wv^{}_{RL}$ [Eq. \eqref{eq:WRLWLR}], respectively,
are completely determined by the AB and SOI phases. To see this,
note that the eigenvectors of $\Wv^{}_{RL}\Wv^{}_{LR}$ and
$\Wv^{}_{LR}\Wv^{}_{RL}$ are simply the eigenvectors of
$u+u^{\dagger}$ and $u'+u'^{\dagger}$, respectively, where
$u=U^{\dagger}_{b}U^{}_{c}=U^{-M}_{Lb}U^{-M}_{bR}U^{M}_{cR}U^{M}_{Lc}$
and
$u'=U^{}_{b}U^{\dagger}_{c}=U^{M}_{bR}U^{M}_{Lb}U^{-M}_{Lc}U^{-M}_{cR}$.
Since the hopping matrices $U^{}_{uv}$ depend only on the AB and
SOI phases, the directions $\pm\hat{\nv}$ and $\pm\hat{\nv}'$ are
determined solely by these phases and are not affected by the
current leakage. Hence, the blocked and transmitted directions can
be manipulated by the external electric and magnetic fields.

From Eq. \eqref{eq:WRLWRL eigenvalues} it follows that
$\lambda^{}_{LR,\pm}\geq0$ and the equality $\lambda^{}_{LR,-}=0$
holds only if
\begin{align}
\label{eq:filtering conditions}
&|\gamma^{}_{b}|=|\gamma^{}_{c}|\equiv\gamma, \nonumber\\
&\cos\left(\widetilde{\phi}-\omega\right)=-1.
\end{align}
We now discuss these two conditions.

The first condition in Eqs. \eqref{eq:filtering conditions} can be
interpreted as a requirement for a symmetry relation between the
two paths. Recall that $\gamma^{}_{b}$ and $\gamma^{}_{c}$ are the
complex effective hopping amplitudes for the two branches of the
loop. In the case of the lossless diamond, we required that this
condition be satisfied independently of the electron energy
$\epsilon$. In the present case, this symmetry can be achieved  by
tuning the various parameters of edges $Lb$ and $bR$ to be equal
to those of edges $Lc$ and $cR$, i.e.\ either
\begin{align}
\label{eq:symmetry relations1}
&J^{}_{Lb}=J^{}_{Lc}, \qquad J^{}_{bR}=J^{}_{cR}, \nonumber\\
&J^{}_{x,Lb}=J^{}_{x,Lc}, \qquad J^{}_{x,bR}=J^{}_{x,cR},
\end{align}
or
\begin{align}
\label{eq:symmetry relations2}
&J^{}_{Lb}=J^{}_{cR}, \qquad J^{}_{bR}=J^{}_{Lc}, \nonumber\\
&J^{}_{x,Lb}=J^{}_{x,cR}, \qquad J^{}_{x,bR}=J^{}_{x,Lc},
\end{align}
and in addition $\epsilon^{}_{0,b}=\epsilon^{}_{0,c}$.  However,
it is not obvious that such a tuning of the parameters can be
realized in an experimental setup.

Alternatively, one can work in the linear-response regime, where
all the electrons have the same energy, equal to the Fermi energy
of the leads. In this case, the first condition in Eqs.
\eqref{eq:filtering conditions} should be satisfied for a single
specific energy, and this can be achieved by tuning only a single
gate voltage, e.g.\ that which controls the site energy
$\epsilon^{}_{0,b}$ (or $\epsilon^{}_{0,c}$).

The second condition in Eqs. \eqref{eq:filtering conditions},
namely $\omega=\widetilde{\phi}+\pi$, imposes a relation between
the AB flux and the SOI strength. We remind the reader that
$\widetilde{\phi}=\phi+\delta^{}_{c}-\delta^{}_{b}$, with
$\delta^{}_{b}$ and $\delta^{}_{c}$ being the phases of
$\gamma^{}_{b}$ and $\gamma^{}_{c}$, respectively. This should be
compared with the condition $\omega=\phi+\pi$, derived for the
lossless diamond.\cite{AA11} As emphasized in Ref.
\onlinecite{AA11}, this relation for the lossless diamond depends
neither on the electron energy $\epsilon$, nor on the site
energies, since $\phi$ and $\omega$ depend only on the unitary
matrices $U^{}_{i}$. In the lossy diamond, on the other hand, the
relation $\omega=\widetilde{\phi}+\pi$ generally depends both on
the electron energy and on the site energies [since
$\gamma^{}_{b}$ and $\gamma^{}_{c}$ depend on these parameters,
see Eqs. \eqref{eq:gamma}]. However, with one of the choices
\eqref{eq:symmetry relations1} or \eqref{eq:symmetry relations2},
the two paths are completely symmetric, so that
$\delta^{}_{b}$=$\delta^{}_{c}$ and the relation between the AB
flux and the SOI strength becomes identical to that of the
lossless diamond. Alternatively, working in the linear-response
regime, one has to tune experimentally the AB flux to satisfy the
condition $\omega=\phi+\delta^{}_{c}-\delta^{}_{b}+\pi$ for
specific values of $\delta^{}_{b}$ and $\delta^{}_{c}$ (which are
determined by the Fermi energy of the leads and by the various
hopping amplitudes, site energies and leakage parameters). For
further details, see the appendix.

Finally, we comment on the magnitude of the transmission of the
polarized electrons through the filter. Since one does not expect
the tight-binding model to be valid near the band edges, we
confine ourselves to the center of the band, $\epsilon=0$ or
$ka=\pi/2$, where the details of the model chosen are not so
important. In the lossless diamond, it was shown in Ref.
\onlinecite{AA11} that by fixing $j$ and $J^{}_{uv}\equiv J$, one
has $T^{}_{+}(\epsilon=0)=|t^{}_{+}(\epsilon=0)|^{2}=1$ at
$\phi=\phi^{}_{0}$ if the various parameters are tuned to be
$\gamma=\gamma^{}_{0}=j/\left(2\sin\phi^{}_{0}\right)$,
$\epsilon^{}_{0,b}=-J^{2}/\gamma^{}_{0}$, and
$\epsilon^{}_{0,L}=\epsilon^{}_{0,R}=-2\gamma^{}_{0}$. Using Eq.
\eqref{eq:transmission amplitudes2} with these choices, the
transmission $T^{}_{+}(\epsilon=0)=|t^{}_{+}(\epsilon=0)|^{2}$
takes the form\cite{AA11}
\begin{align}
\label{eq:transmission no losses}
&T^{}_{+}(\epsilon=0,\phi)=\frac{4\sin^{2}\phi\sin^{2}\phi^{}_{0}}{\left(\sin^{2}\phi+\sin^{2}\phi^{}_{0}\right)^{2}}.
\end{align}
This function is plotted in Fig. \ref{fig:transmission vs flux}
for two values of $\phi^{}_{0}$ and for $J=4j$. To compare with
the transmission $T^{}_{+}$ of the lossy diamond, we use the same
parameters chosen above, but now we "turn on" the leakage by
setting $M=5$ and $J^{}_{x,uv}=0.2j$. The results are shown in
Fig. \ref{fig:transmission vs flux}. Figure \ref{fig:transmission
vs k} shows the transmission $T^{}_{+}$ as a function of $ka$ with
the AB flux fixed at $\phi=\phi^{}_{0}$, for the parameters chosen
above and for $J=4j$. Two curves correspond to the lossless
diamond ($M=1$ and $J^{}_{x,uv}=0$) and the other two correspond
to the lossy diamond with $M=5$ and $J^{}_{x,uv}=0.2j$.
\begin{figure}[ht]
\begin{center}
\subfigure[\label{fig:transmission vs flux}]{
\label{fig:transmission vs flux}
\includegraphics[width=0.52\textwidth,height=0.24\textheight]{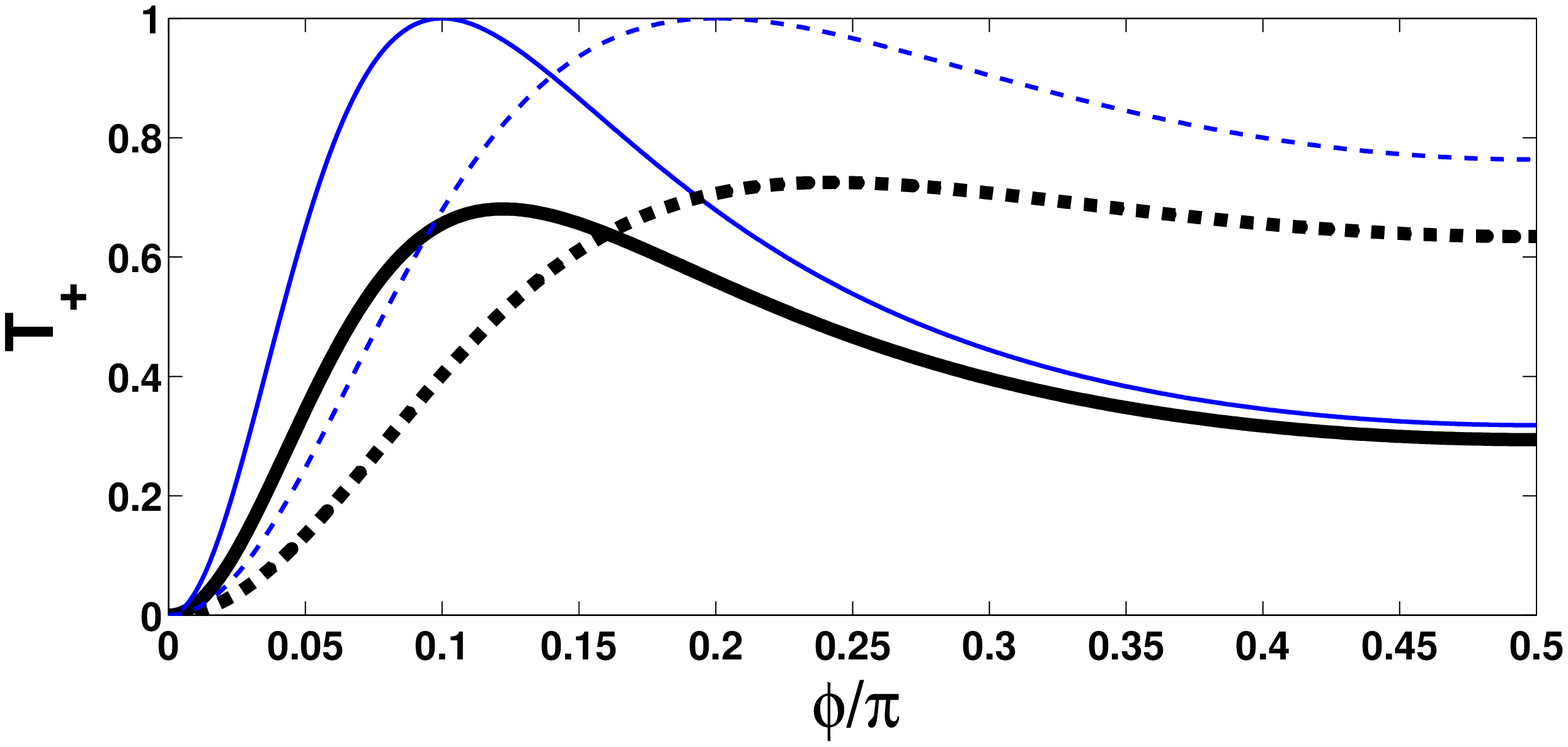}
} \subfigure[\label{fig:transmission vs k}]{
\label{fig:transmission vs k}
\includegraphics[width=0.52\textwidth,height=0.24\textheight]{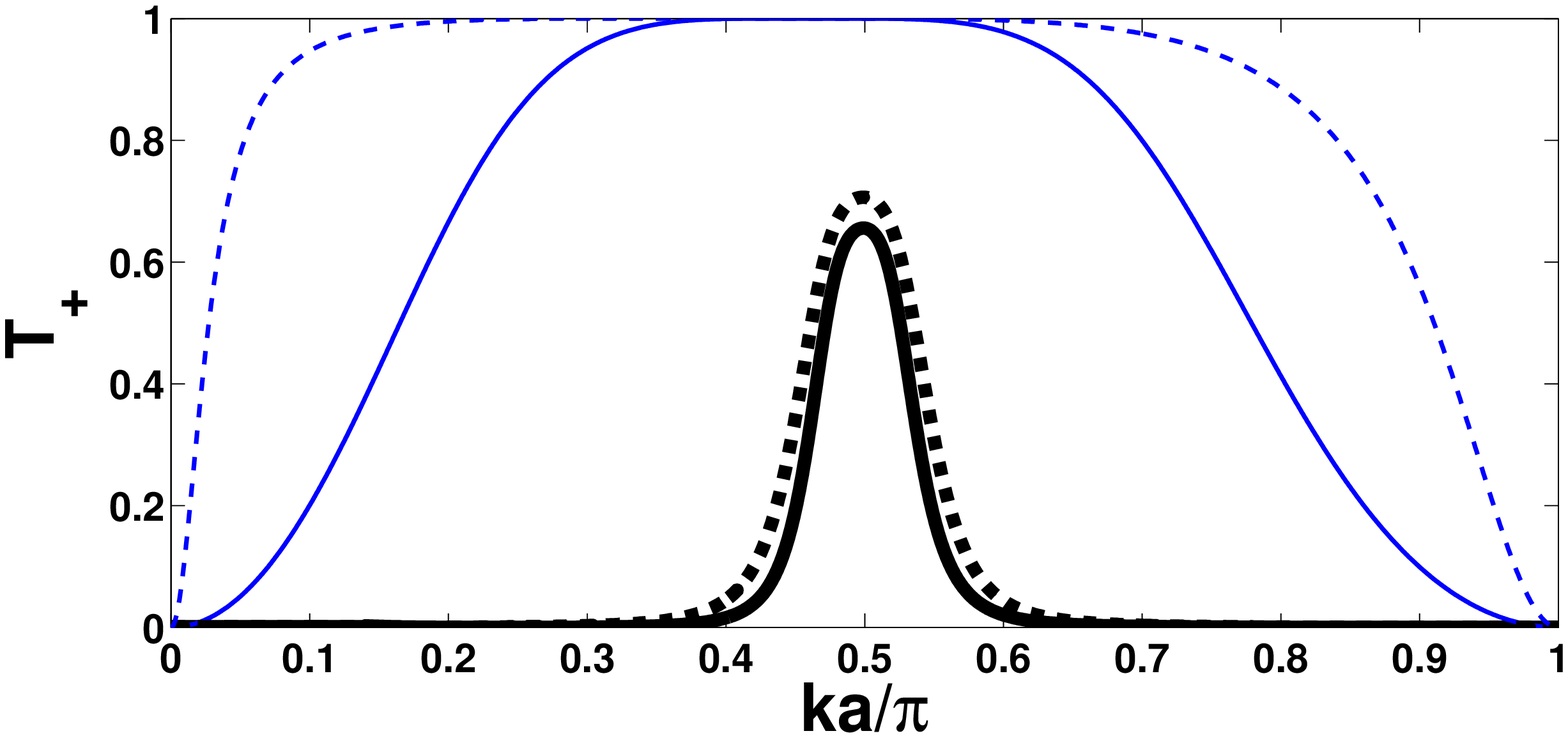}
}
\end{center}
\caption{\label{fig:transmission}(Color online) The transmission
of the polarized electrons, $T^{}_{+}(\epsilon,\phi)$ (a) as a
function of the AB flux $\phi$ (in units of $\pi$) for
$\epsilon=0$ ($ka=\pi/2$) and (b) as a function of $ka$ (in units
of $\pi$) for $\phi=\phi^{}_{0}$. Solid and dashed curves
correspond to maxima of $T^{}_{+}(\epsilon=0,\phi)$ at
$\phi^{}_{0}=0.1\pi$ and $\phi^{}_{0}=0.2\pi$, respectively. Thin
(blue) and thick (black) curves correspond to the lossless ($M=1$
and $J^{}_{x,uv}=0$) and lossy ($M=5$ and $J^{}_{x,uv}=0.2j$)
diamonds, respectively.}
\end{figure}
Comparing the transmissions of lossless and the lossy diamonds
presented in Fig. \ref{fig:transmission}, several differences can
be detected. First, the maximal transmission of the lossy diamond
is less than one. This is an obvious effect of the lossy diamond,
since part of the incident current leaks into the absorbing
channels. This leakage increases as the tunnelling from the
diamond edges to each absorbing channel, represented by
$J^{}_{x,uv}$, increases or when the number of absorbing channels
$M$ increases. Second, Fig. \ref{fig:transmission vs flux} shows
that the maximal transmission is not obtained for
$\phi=\phi^{}_{0}$ but for a flux slightly larger than
$\phi^{}_{0}$. Third, in Fig. \ref{fig:transmission vs k} we see
that the transmission as a function of $ka$ is much narrower than
that of the lossless case.
\section{Summary and Conclusions}
\label{Summary} We have generalized the results of the
single-diamond interferometer made of materials with significant
SOIs and penetrated by an AB flux, discussed in Ref.
\onlinecite{AA11}, to the case where losses are present. In
particular, we have considered losses due to a current leakage out
of the diamond into absorbing channels, modelled as 1D
tight-binding chains. Our calculations show that spin filtering
(and consequently also spin reading) can be achieved even in the
presence of current leakage. The filtering conditions found in
Ref. \onlinecite{AA11}, namely the condition for a symmetry
relation between the two branches
$\left(\gamma^{}_{b}=\gamma^{}_{c}\right)$ and the relation
between the AB phase and the SOI phase
$\left(\phi=\omega+\pi\right)$ are modified. In the lossy diamond
$\gamma^{}_{b}$ and $\gamma^{}_{c}$ are complex numbers and the
first condition takes the form $|\gamma^{}_{b}|=|\gamma^{}_{c}|$.
Since $\gamma^{}_{b}$ and $\gamma^{}_{c}$ depend on the various
parameters (leakage parameters $J^{}_{x,uv}$, hopping amplitudes
$J^{}_{uv}$, electron energy $\epsilon$ and site energies
$\epsilon^{}_{0,b}$ and $\epsilon^{}_{0,c}$) in a complicated
manner, it is difficult to derive general analytical relations
between these parameters from this condition. The solution
simplifies for the completely symmetric diamond [Eqs.
\eqref{eq:symmetry relations1} and \eqref{eq:symmetry relations2}]
for which the condition $|\gamma^{}_{b}|=|\gamma^{}_{c}|$ is
trivially satisfied. The second condition, $\phi=\omega+\pi$, is
still valid provided that $\phi$ is replaced by
$\widetilde{\phi}=\phi+\delta^{}_{c}-\delta^{}_{b}$, with
$\delta^{}_{b}$ and $\delta^{}_{c}$ being the phases of
$\gamma^{}_{b}$ and $\gamma^{}_{c}$, respectively. For the
completely symmetric diamond one has $\delta^{}_{b}=\delta^{}_{c}$
and this condition is the same as for the lossless diamond. An
alternative way to satisfy the filtering conditions is by working
in the linear-response regime. In that regime, transport of
electrons occurs at the Fermi energy of the leads. The filtering
conditions can then be satisfied simultaneously by tuning the AB
flux and one of the site energies $\epsilon^{}_{o,b}$ or
$\epsilon^{}_{o,c}$.

Once the filtering conditions are obeyed, the transmitted
electrons are fully polarized. The effects of the leakage on the
transmission can be summarized as follows:
\begin{enumerate}[(1)]
\item The maximal transmission in the lossy diamond is always
smaller than one and the maximum decreases with increasing leakage
parameters or with increasing length of each edge of the diamond.
\item Viewed as a function of the flux $\phi$, the transmission of
the lossy diamond is shifted to slightly higher fluxes relative to
the transmission of the lossless diamond, while viewed as a
function of $ka$, the transmission of the lossy diamond is
narrower than the transmission of the lossless diamond.
\end{enumerate}
However, it should be emphasized that quantities which depend only
on the hopping matrices $U^{M}_{uv}$, are not affected by the
leakage. For instance, the dependence of the SOI phase $\omega$ on
the SOI strength and on the geometry of the diamond (through the
opening angle $2\beta$) is the same as in the lossless diamond and
so are the directions $\hat{\nv}$ and $\hat{\nv}'$ of the filtered
and the transmitted electrons. Analytical expressions for the SOI
phase and for the blocked and transmitted spinors in the lossless
diamond have been obtained in Ref. \onlinecite{AA11} for the
Rashba-only SOI and for both Rashba and Dresselhaus
SOI.\cite{comment3} Those expressions thus remain valid for the
lossy diamond.

In conclusion, while the filtering conditions are slightly
modified, many of the device properties remain the same as in the
lossless case. These properties can be manipulated by external
electric and magnetic fields and are insensitive to the current
leakage in the framework of the model presented here.
\section*{Appendix: spin filtering in the asymmetric interferometer}
\label{Summary}  
\begin{figure}[ht]
\begin{center}
\subfigure[\label{fig:gamma abs}]{ \label{fig:gamma abs}
\includegraphics[width=0.5\textwidth,height=0.25\textheight]{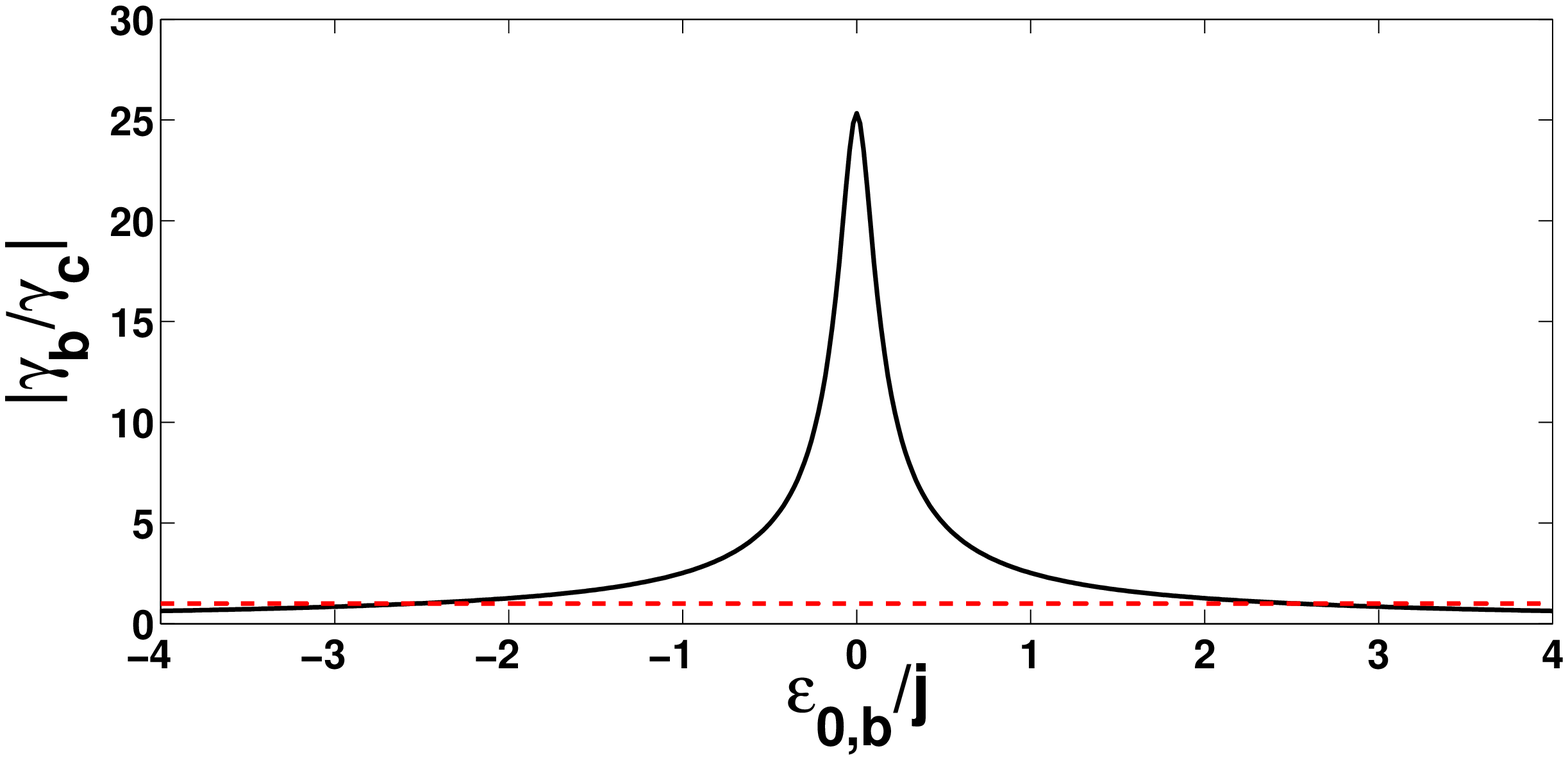}
} \subfigure[\label{fig:gamma phase}]{ \label{fig:gamma phase}
\includegraphics[width=0.5\textwidth,height=0.25\textheight]{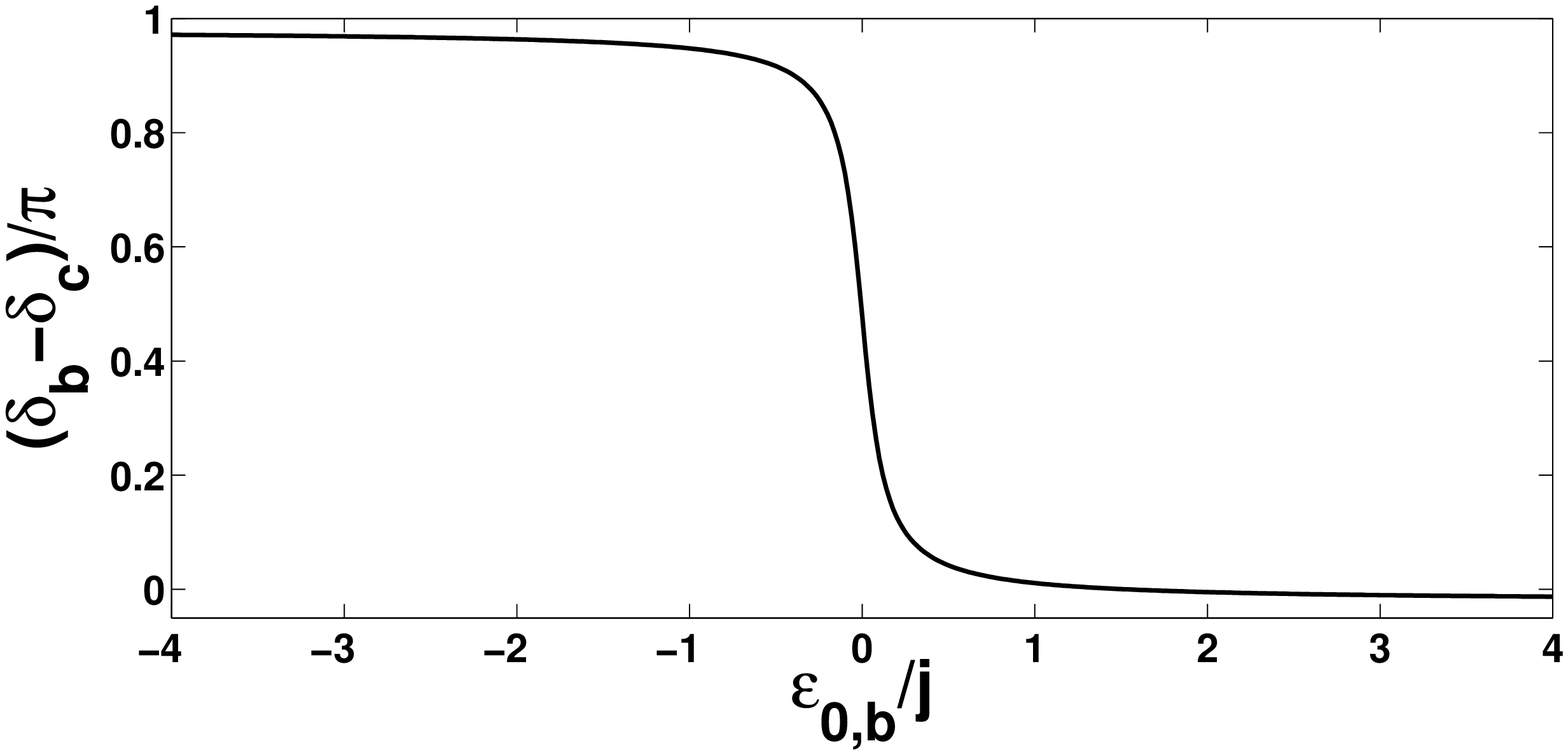}
}
\end{center}
\caption{\label{fig:gammas}(Color online) The dependence of (a)
$|\gamma^{}_{b}/\gamma^{}_{c}|$ and (b)
$\delta^{}_{b}-\delta^{}_{c}$ (in units of $\pi$) on the site
energy $\varepsilon^{}_{0,b}$ (in units of $j$) for
$J^{}_{Lb}=3j$, $J^{}_{bR}=4j$, $J^{}_{Lc}=2.5j$,
$J^{}_{cR}=3.8j$, $J^{}_{x,Lb}=0.1j$, $J^{}_{x,bR}=0.2j$,
$J^{}_{x,Lc}=0.05j$, $J^{}_{x,cR}=0.25j$, $M=5$ and
$\epsilon^{}_{0,c}=2j$. The dashed (red) line in (a) corresponds
to $|\gamma^{}_{b}/\gamma^{}_{c}|=1$.}
\end{figure}

We have pointed out that realizing a symmetric
interferometer in the experiment may be a difficult task. We have
mentioned that the filtering conditions \eqref{eq:filtering
conditions} can be fulfilled for an arbitrary asymmetric
interferometer by working in the linear-response regime. Let us
assume that the Fermi energy of the leads lies at the center of
the tight-binding band (i.e.\ $\epsilon=\epsilon^{}_{F}=0$ or
$ka=\pi/2$).  Then, one can satisfy the first condition in Eqs.
\eqref{eq:filtering conditions} (with $\epsilon=0$) by tuning only
the site energy $\epsilon^{}_{0,b}$. The phase
$\delta^{}_{b}-\delta^{}_{c}$ is then fixed, and one can satisfy
the second condition in Eqs. \eqref{eq:filtering conditions} by
tuning the AB flux. To examine the variation of
$\gamma^{}_{b}/\gamma^{}_{c}$ as a function of
$\epsilon^{}_{0,b}$, we set $J^{}_{Lb}=3j$, $J^{}_{bR}=4j$,
$J^{}_{Lc}=2.5j$, $J^{}_{cR}=3.8j$, $J^{}_{x,Lb}=0.1j$,
$J^{}_{x,bR}=0.2j$, $J^{}_{x,Lc}=0.05j$, $J^{}_{x,cR}=0.25j$,
$M=5$ and $\epsilon^{}_{0,c}=2j$ and plot
$|\gamma^{}_{b}/\gamma^{}_{c}|$ and $\delta^{}_{b}-\delta^{}_{c}$
in Fig. \ref{fig:gammas} as a function of $\epsilon^{}_{0,b}$. The
dashed (red) line in Fig. \ref{fig:gamma abs} corresponds to
$|\gamma^{}_{b}/\gamma^{}_{c}|=1$. The intersection of the two
curves occurs at the values $\epsilon^{*}_{0,b}$ for which the
first condition in Eqs. \eqref{eq:filtering conditions} holds.
Figure \ref{fig:gamma abs} shows that
$|\gamma^{}_{b}/\gamma^{}_{c}|$ changes slowly at the vicinity of
$\epsilon^{*}_{0,b}$. Furthermore, Fig. \ref{fig:gamma phase}
shows that the phase $\delta^{}_{b}-\delta^{}_{c}$ is also a
slowly varying function of $\epsilon^{}_{0,b}$ at the vicinity of
$\epsilon^{*}_{0,b}$. This means that the fulfillment of the
filtering conditions \eqref{eq:filtering conditions} is not
sensitive to small deviations of $\epsilon^{}_{0,b}$ from
$\epsilon^{*}_{0,b}$. The results presented in Fig.
\ref{fig:gammas} have a weak dependence on the various energies.
The width and the height of the peak in Fig. \ref{fig:gamma abs}
and the width of the crossover region in Fig. \ref{fig:gamma
phase} change slightly with the various energies, but the overall
shape is robust.
\begin{acknowledgments}
We acknowledge support from the Israel Science Foundation (ISF).
\end{acknowledgments}


\begin{references}
\bibitem{comment1} Below we refer to $\phi$ as the AB flux in units of $\Phi^{}_{0}/2\pi$.
\bibitem{comment2} We note that for $M=1$, in which case there is no leakage to the absorbing channels, Eqs. \eqref{eq:effective_hopping1} and \eqref{eq:effective_hopping2} reduce to $J'^{}_{uv}=0$ and
$J''^{}_{uv}=J^{}_{uv}$, respectively. In that case Eqs.
\eqref{eq:corners equations2} are identical to Eqs. (7) of Ref.
\onlinecite{AA11}.
\bibitem{comment3} Given the spin-orbit interactions on the four edges of the diamond, one computes the SOI phase $\omega$ as follows. First one calculates the matrices $U^{M}_{uv}$. As discussed in the introduction, these matrices are
of the form
$U^{M}_{uv}=e^{i\phi^{}_{uv}+i\Kv^{}_{uv}\cdot\sigmav}$ with
$\Kv^{}_{uv}$ given by Eq. \eqref{eq:Rotation matrix1} and
$\phi^{}_{uv}$ being the AB phase assigned to the edge $uv$ (which
depends on the gauge of the vector potential $\Av$). From these
matrices one calculates explicitly the matrix
$u=U^{\dagger}_{b}U^{}_{c}=U^{-M}_{Lb}U^{-M}_{bR}U^{M}_{cR}U^{M}_{Lc}$.
Comparing with the form
$u=e^{i\phi+i\omegav\cdot\sigmav}=e^{i\phi}\left(\cos\omega+i\sin\omega\hat{\omegav}\cdot\sigmav\right)$,
one extracts the relations for $\cos\omega$, $\sin\omega$ and the
blocked direction $\hat{\nv}=-\hat{\omegav}$.
\end{references}
\end{document}